\documentclass[aps,prstab,reprint,groupedaddress,showpacs,amsmath,amssymb]{revtex4-1}
\usepackage{epsfig}
\usepackage{dcolumn}
\usepackage{bm}

\begin{document}


\title{Preparation of an emittance transfer experiment}


\author{C.~Xiao and O.~Kester}
\affiliation{Institut Angewandte Physik, Goethe-Universit\"at Frankfurt, Frankfurt am Main D60438, Germany}
\author{L.~Groening, H.~Leibrock, and M.~Maier}
\affiliation{GSI Helmholtzzentrum f\"ur Schwerionenforschung GmbH, Darmstadt D64291, Germany}


\date{\today}

\begin{abstract}
Flat beams feature unequal emittances in the horizontal and vertical phase space.
Those beams were created successfully in lepton machines. Although a number of applications will profit also from flat hadron beams,
to our knowledge they have never been created systematically. Multi-turn injection schemes, spectrometers, and colliders will
directly benefit from those beams. The present paper covers the preparation of the experimental proof of principle for
flat hadron beam creation in a beam transport section. Detailed simulations of the experiment, based on charge state stripping inside of a solenoid [L. Groening, Phys. Rev. ST Accel. Beams 14, 064201 (2011)], are performed. The matrix formalism was benchmarked with tracking through three-dimensional magnetic field maps of solenoids. An error analysis targeting at investigation of the impact of machine errors on the round-to-flat beam
transformation has been performed. The remarkable flexibility of the set-up w.r.t. decoupling is addressed, as it can provide an one-knob tool to set the horizontal and vertical emittance partitioning. Finally, the status of hardware design and production is given.
\end{abstract}


\maketitle


\section{Introduction}
The modification of projected beam emittances under preservation of the full
six-dimensional emittance became a matter of interest for many accelerator
applications.
First experiments were proposed and conducted by D.~Edwards et al.~\cite{Edwards} for
electron machines about a decade ago. The issue is of
special interest for increasing the performance of X-FELs and advanced approaches
to emittance repartitioning are under conceptual and experimental
investigation~\cite{Carlsten_May2011,Carlsten_Aug2011,Piot,Sun,Xiang}. Flat hadron beams
could facilitate the process of multi-turn injection into circular machines, which
imposes different requirements on the horizontal and vertical emittance of the
incoming beam. Recently it was proposed to use flat beams in hadron-hadron collisions
to provide higher luminosity by mitigating beam-beam effects~\cite{Burov}.
The mass resolution of spectrometers is increased significantly if the beam is flat
perpendicular to the direction of the spectrometers bend. A corresponding set-up behind an
Electron-Cyclotron-Resonance source is proposed in~\cite{Bertrand}.
\\
From first principles beams are created round without any coupling among planes.
Their rms emittances as well as their eigen-emittances are equal in the two transverse planes. Thus, any transverse round-to-flat transformation requires a change of the beam eigen-emittances by a non-symplectic transformation~\cite{Dragt}. Such a transformation can be performed by placing a charge state stripper inside an axial magnetic field region as proposed in~\cite{Groening}. Inside such a solenoid stripper, transverse inter-plane correlations are created non-symplectically. Afterwards they are removed symplectically by a decoupling section including skew quadrupoles. It must be mentioned that the use of charge state strippers (outside from solenoids) is state-of-the art at several ion machines that provide highly charged ions.
\\
It is emphasized that the paper is on the application of coupled beam dynamics aiming for increased performance of
an accelerator chain. It is not on coupled beam dynamics theory itself and references are given whenever needed.
The paper starts with a re-introduction of the required terms of coupled beam
dynamics. Afterwards the new set-up for experimental demonstration of transverse emittance transfer is introduced. The fourth section is on modeling the non-symplectic process of
charge state stripping inside a solenoid. Results from models
based on matrix formalism are benchmarked with those from tracking particles through
three-dimensional field maps derived from magnet design codes. Such benchmarks were made
for different finite fringe field shapes of the solenoid. Afterwards, the symmplectic
decoupling section is treated. The sixth section analyses the impact of machine errors
on the decoupling performance of the beam line. It was found that the decoupling
capability of the set-up is remarkably flexible and the impact and discussion of this
finding is treated in a dedicated section. The paper closes with some conclusions and
an outlook w.r.t. procurement of the required hardware.

\section{Basic terms}
The four-dimensional symmetric beam matrix $C$ contains ten unique elements, four of which describe the coupling. If one or more of the elements of the off-diagonal sub-matrix is non-zero, the beam is $x$-$y$ coupled:
\begin{equation}
C=
\begin{bmatrix}
\langle xx \rangle &  \langle xx'\rangle &  \langle xy\rangle & \langle xy'\rangle \\
\langle x'x\rangle &  \langle x'x'\rangle & \langle x'y\rangle & \langle x'y'\rangle \\
\langle yx\rangle &  \langle yx'\rangle &  \langle yy\rangle & \langle yy'\rangle \\
\langle y'x\rangle &  \langle y'x'\rangle & \langle y'y\rangle & \langle y'y'\rangle
\end{bmatrix}.
\end{equation}
The four-dimensional rms emittance $\varepsilon_{4d}$ is the square root of the determinant of $C$, and the projected beam rms emittances $\varepsilon_x$ and $\varepsilon_y$ are the square roots of the determinants of the on-diagonal sub-matrices. Diagonalization of the beam matrix yields the eigen-emittances $\varepsilon_1$ and $\varepsilon_2$ which are calculated as
\begin{align}
\label{eq2}
{\varepsilon_{1,2}}=\frac{1}{2} \sqrt{-tr(CJ)^2 \pm \sqrt{tr^2(CJ)^2-16 det(C) }}.
\end{align}
The four-dimensional matrix $J$ is the skew-symmetric matrix with non-zero entries on the block diagonal off form. Any symplectic transformation $M$ obeys
\begin{equation}
M^TJM=J,
~~~J=
\begin{bmatrix}
0 &  1 &  0 & 0 \\
-1 &  0 &  0 & 0 \\
0 &  0 &  0 & 1 \\
0 &  0 & -1 & 0
\end{bmatrix}.
\end{equation}
Eigen-emittances are invariant under symplectic transformations and the eigen-emittances are equal to the rms emittances only if inter-plane ($x$-$y$) correlations are zero.

\section{Experimental set-up}
\begin{figure}[hbt]
\centering
\includegraphics*[width=80mm,clip=]{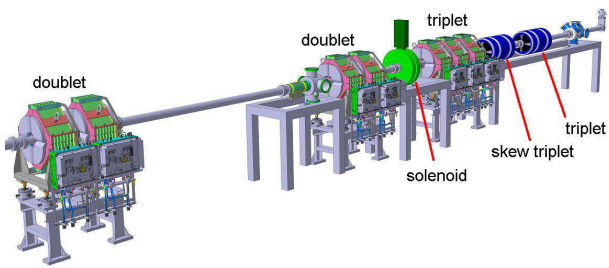}
\caption{The layout of the EMTEX section at GSI.}
\label{fig_beamline}
\end{figure}

The new EMTEX (emittance transfer experiment) beam line for the demonstration of transverse emittance transfer is shown in~Fig.~\ref{fig_beamline} and it will be integrated into the existing transfer line from the UNILAC~\cite{Barth} to the SIS-18 synchrotron.
\\
\\
The transverse emittance transfer beam line comprises two quadrupole doublets, a solenoid with stripper foil inside, a quadrupole triplet, a skew quadrupole triplet, another quadrupole triplet, a current transformer, and a transverse emittance measurement unit. Its total length is 12785~mm.
\\
In order to mitigate four-dimensional rms emittance growth from scattering during the
stripping process, the beam sizes at the stripper should be kept small. Two quadrupole doublets separated by a drift space in front of the solenoid do the required matching. The maximum gradients of the quadrupole magnets are 19.0 and 15.0~T/m and the effective field lengths are 319 and 354~mm, respectively.
A low intensity beam of $D_6^+$ stripped to 3$D_2^+$ in a 20~$\mu g/cm^2$ carbon foil placed at the center of a solenoid will be used, and the total relative momentum spread of the beam is less than $\pm5 \times 10^{-4}$. The maximum longitudinal magnetic field is 1.0~T. This non-symplectic transformation creates coupling between the two transverse planes~\cite{Non-symplecticity}. A quadrupole triplet and a skew quadrupole triplet separated by a drift space are employed to remove these correlation symplectically.
It will be called decoupling section in the following. A final quadrupole triplet is used for matching to the existing beam line followed by a beam current transformer and an
emittance measurement unit. The full beam line is presented quantitatively in the Appendix.

\section{stripping inside a solenoid}

\begin{figure}[hbt]
\centering
\includegraphics*[width=80mm,clip=]{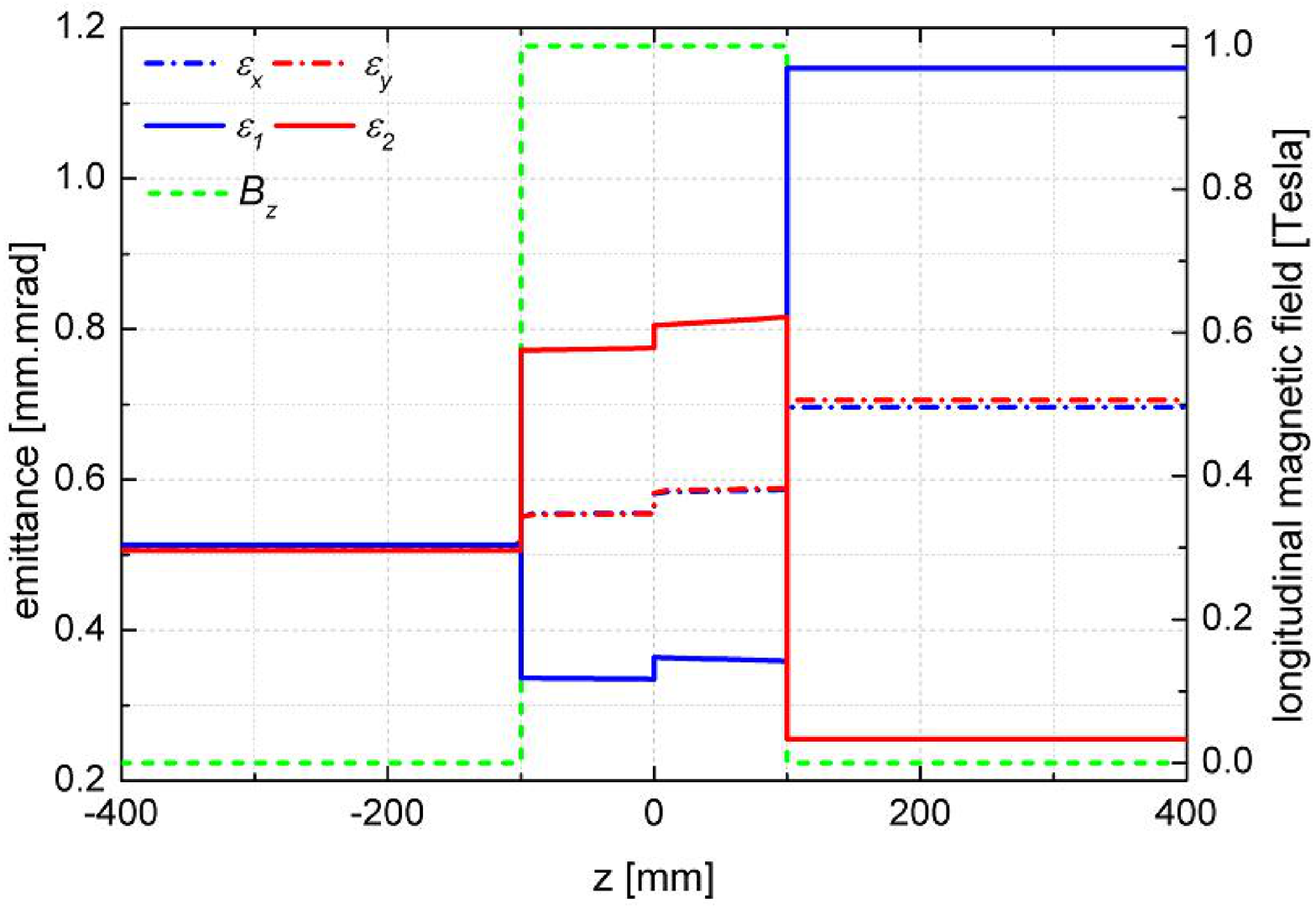}
\caption{Evolutions of rms emittances and eigen-emittances along the solenoid channel using a transfer matrix.}
\label{matrix1}
\end{figure}

\begin{figure}[hbt]
\centering
\includegraphics*[width=80mm,clip=]{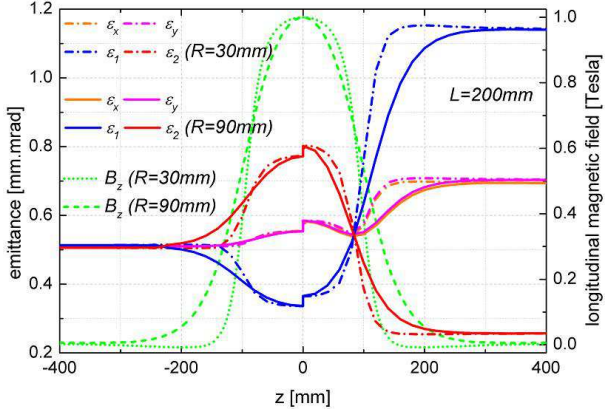}
\caption{Evolutions of rms emittances and eigen-emittances along the solenoid, with same effective lengths but different aperture radii. The results are from tracking through three-dimensional magnetic field maps.}
\label{1D field 1}
\end{figure}

\begin{figure}[hbt]
\centering
\includegraphics*[width=80mm,clip=]{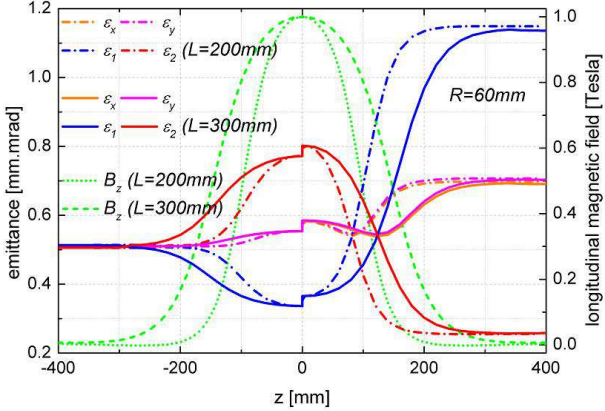}
\caption{Evolutions of rms emittances and eigen-emittances along the solenoid, with identical inner radius but different effective lengths. The results are from tracking through three-dimensional magnetic field maps.}
\label{1D field 2}
\end{figure}

Stripping inside a solenoid is fundamentally different from stripping between two solenoids due to the longitudinal magnetic field component and the fringe fields. In case of pure transverse field components (dipoles, quadrupoles, n-poles) there is equivalence between stripping inside this magnet and stripping between two such magnets of half lengths.
\\
Let $C_0$ denote the second moment matrix at the entrance of the solenoid. If the beam has equal horizontal and vertical rms emittances, the beam matrix can be simplified to
\begin{equation}
C_0=
\begin{bmatrix}
\varepsilon \beta &  0 &  0 & 0 \\
0 &  \frac{\varepsilon} {\beta} &  0 & 0 \\
0 &  0 &  \varepsilon \beta & 0 \\
0 &  0 & 0 & \frac{\varepsilon} {\beta}
\end{bmatrix}\,.
\end{equation}
Assuming a very short solenoid, its transport matrix can be divided into two parts
\begin{equation}
R_{in}=
\begin{bmatrix}
1 &  0 &  0 & 0 \\
0 &  1 &  k_{in} & 0 \\
0 &  0 &  1 & 0 \\
-k_{in} &  0 & 0 & 1
\end{bmatrix}
,
R_{out}=
\begin{bmatrix}
1 &  0 &  0 & 0 \\
0 &  1 &  -k_{out} & 0 \\
0 &  0 &  1 & 0 \\
k_{out} &  0 & 0 & 1
\end{bmatrix}.
\end{equation}
If the beam has the same energy and charge state at the solenoid entrance and exit, $k_{in}$ is equal to $k_{out}$. The first part describes the entrance fringe field and the second part is the exit fringe field. In here, the focusing strength of the solenoid is
\begin{align}
\label{eq2}
{k=\frac{B}{2(B \rho)} }.
\end{align}
$B$ is the on-axis magnetic field strength, and $B\rho$ is the beam rigidity. The beam matrix $C_1$ after the entrance fringe field $k$ is found in the following form
\begin{equation}
C_1=R_{in}C_1R_{in}^T=
\begin{bmatrix}
\varepsilon \beta &  0 &  0 & -k\varepsilon \beta \\
0 &  \frac{\varepsilon} {\beta}+k^2\varepsilon \beta &  k\varepsilon \beta & 0 \\
0 &  k\varepsilon \beta &  \varepsilon \beta & 0 \\
-k\varepsilon \beta &  0 & 0 & \frac{\varepsilon} {\beta}+k^2\varepsilon \beta
\end{bmatrix}.
\end{equation}
The off-diagonal sub-matrices describe the correlations and the values of $\langle xy\rangle$ and $\langle x'y'\rangle$ are zero.
In order to achieve a change of the eigen-emittances a non-symplectic transformation has to be integrated into the round-to-flat transformation section. The transformation through the solenoid is non-symplectic if the beam rigidity is abruptly changed in between the entrance and exit fringe fields, thus the beam properties are reset inside the solenoid. The non-symplectic transformation is accomplished by using a beam of $D_6^+$ stripped to 3$D_2^+$ in a carbon foil placed at the center of solenoid. Thus, the exit fringe field transfer matrix is changed to:
\begin{equation}
R_{out}^{'}=
\begin{bmatrix}
1 &  0 &  0 & 0 \\
0 &  1 &  -3k & 0 \\
0 &  0 &  1 & 0 \\
3k &  0 & 0 & 1
\end{bmatrix}.
\end{equation}
The focusing strength of the solenoid $k$ is calculated from the unstripped charge state. The elements of the beam matrix $C'_1$ directly after the stripper inside of the solenoid but still before the exit fringe field are
\begin{equation}
C_{1}^{'}=
\begin{bmatrix}
\varepsilon \beta &  0 &  0 & -k\varepsilon \beta \\
0 &  \frac{\varepsilon} {\beta}+k^2\varepsilon \beta+\Delta\varphi^2  &  k\varepsilon \beta & 0 \\
0 &  k\varepsilon \beta &  \varepsilon \beta & 0 \\
-k\varepsilon \beta &  0 & 0 & \frac{\varepsilon} {\beta}+k^2\varepsilon \beta+\Delta\varphi^2
\end{bmatrix}\,.
\end{equation}
The stripper scattering effects on the angular spread are included. The energy loss and straggling in the stripper foil can be neglected in our case. The parameter $\Delta\varphi^2$ is the scattering amount during the stripping process~\cite{Atima}, and the foil stripper itself is modeled by increasing the spread of the angular distribution through scattering.
After the stripper the beam passes through the exit fringe field with reduced beam rigidity and the beam matrix $C_{2}^{'}$ after the exit fringe field becomes
\begin{equation}
{C_{2}^{'}=R_{out}^{'}C_{1}^{'} R_{out}^{'T}=}
\begin{bmatrix}
\varepsilon_n R_n & 2k\varepsilon_n \beta_n J_n \\
-2k\varepsilon_n \beta_n J_n & \varepsilon_n R_n
\end{bmatrix}\,,
\end{equation}
where
\begin{align}
\label{eq2}
\varepsilon_n=\sqrt{ \varepsilon \beta (\frac{\varepsilon}{\beta}+4k^2 \varepsilon \beta + \Delta \varphi^2}), ~~~\beta_n=\frac{\beta \varepsilon}{\varepsilon_n}\,,
\end{align}
introducing the 2$\times$2 sub-matrices $R_n$ and $J_n$ as
\begin{equation}
R_n=
\begin{bmatrix}
\beta_n &  0  \\
0 &  \frac{1}{\beta_n}
\end{bmatrix}\,
,~~~
J_n=
\begin{bmatrix}
0 &  1  \\
-1 &  0
\end{bmatrix}.
\end{equation}
The amount of eigen-emittance transfer scales with the longitudinal magnetic field strength and the beam rms sizes on the stripper. Inter-plane correlations are created and the rms emittances and eigen-emittances after the solenoid with stripper foil can be written as:
\begin{align}
\label{eq2}
\varepsilon_{x,y}=\varepsilon_n, ~~~\varepsilon_{1,2}=\varepsilon_n(1\pm 2k\beta_n)\,.
\end{align}
The four-dimensional rms emittance can be written as
\begin{align}
\label{eq2}
\varepsilon_{4d}=\varepsilon_1\varepsilon_2=\varepsilon^2+\varepsilon \beta  \Delta \varphi^2\,.
\end{align}
The value of the four-dimensional rms emittance increase is proportional to the square of beam sizes on the stripper. The increase is purely from scattering in the foil, it is not caused by the shift of beam rigidity inside the longitudinal magnetic field. The evolutions of the rms emittances and the eigen-emittances along a solenoid channel which is composed of two drift space separated by a solenoid are shown in~Fig.~\ref{matrix1}. For the beam transport through the solenoid, the linear solenoid transfer matrix is used and the stripper foil is placed at the solenoid center. Once the beam feels the entrance fringe, the eigen-emittances start to split out rapidly and do not change until the stripper foil. Taking into account the scattering in the stripper foil, the eigen-emittances increase abruptly during the stripping process. After stripping, the exit fringe is passed by the beam with reduced rigidity, thus overcompensating the previous eigen-emittance variations.
The fringe fields of the solenoid, rather than the pure longitudinal magnetic field, cause the change of eigen-emittances. 
\\
Multi-particle tracking in the proposed transverse emittance transfer section is done with the
TRACK code~\cite{Peter}.
\\
\\
By variations of the aperture size of solenoid, the effective length of the solenoid is constant and the transfer matrix will remain unchanged, while the actual three-dimensional field map will be affected. In order to verify whether usage of the transfer matrix formalism is justified, it was
compared with tracking simulations for two different solenoids of equal effective field length but with different fringe field shapes obtained from three-dimensional field maps of the OPERA-3D finite element code~\cite{OPERA-3D}.
The evolutions of rms emittances and eigen-emittances along solenoids with different aperture radii are shown in~Fig.~\ref{1D field 1}. The solenoids effective lengths are set to 200~mm, and the solenoids aperture radii are chosen to be 30 and 90~mm, respectively.
\\
Additionally we treated the case of fixed solenoid aperture radius but different effective field lengths. The evolutions of the rms emittances and eigen-emittances along solenoids with different effective field
lengths are shown in~Fig.~\ref{1D field 2}.  In this simulation, the solenoid aperture radius is set to 60~mm, and the solenoid effective lengths are chosen 200 and 300~mm, respectively.
\\
The simulations demonstrated that the treatment of the solenoidal stripping process
with linear matrices as initially done in~\cite{Groening} is justified. The final
rms emittances and eigen-emittances do not depend on the exact shape of the fringe
field as long as it is reasonably short like for the solenoids that are commonly in
use. Additional material on this issue can be found in~\cite{Xiao_IAPRep,Xiao_HB2012}.

\section{Decoupling section}
\begin{figure}[hbt]
\centering
\includegraphics*[width=80mm,clip=]{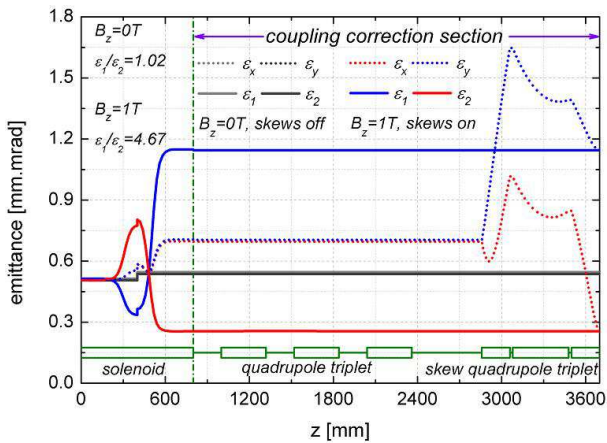}
\caption{Evolution of rms emittances and eigen-emittances along the longitudinal magnetic field and the decoupling section for two scenarios: solenoid and skew quads off (reference, green and dark green lines); solenoid and skew quads on (emittance transfer), $B_z$=1.00~T.}
\label{emittance transfer 1}
\end{figure}

\begin{figure}[hbt]
\centering
\includegraphics*[width=80mm,clip=]{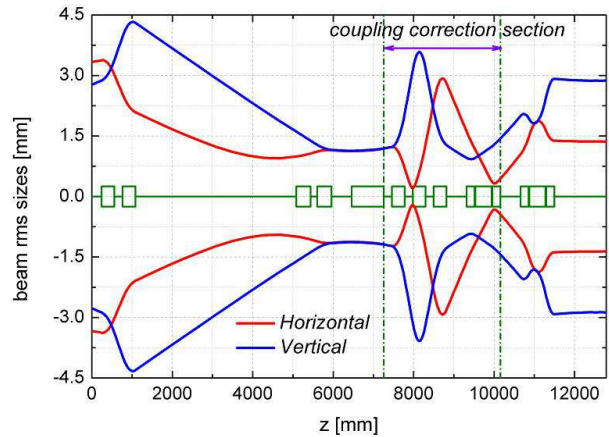}
\caption{Horizontal and vertical beam rms sizes along the proposed transverse emittance transfer section.}
\label{beam rms sizes}
\end{figure}

\begin{figure}[hbt]
\centering
\includegraphics*[width=80mm,clip=]{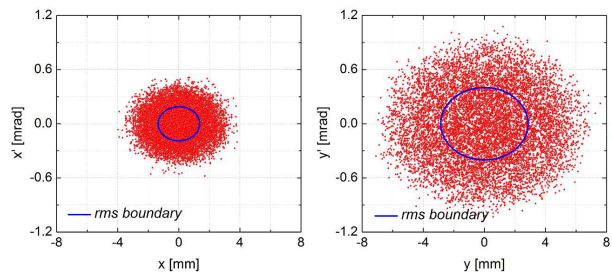}
\caption{The transverse emittance portraits at the exit of proposed transverse emittance transfer section.}
\label{particle distribution}
\end{figure}

The simplest skew decouplig section contains three skew quadrupoles with appropriate betatron phase advances in each plane~\cite{Burov_prstab094002,Burov_pre016503}.
Let $R_q$ be the 4$\times$4 matrix corresponding to a certain arrangement of quadrupoles and drift spaces and assume that this channel is represented by an identity matrix in the $x$-direction and has an additional 90$^\circ$ phase advance in $y$-direction as in~\cite{Kim}
\begin{equation}
R_q=
\begin{bmatrix}
I_n &  O_n  \\
O_n &  T_n
\end{bmatrix}\,.
\end{equation}
Here the 2$\times$2 sub-matrices $O_n$, $T_n$ and $I_n$ are defined as
\begin{equation}
O_n=
\begin{bmatrix}
0 &  0  \\
0 &  0
\end{bmatrix}
,~~~
T_n=
\begin{bmatrix}
0 &  U  \\
-\frac{1}{U} &  0
\end{bmatrix}
,~~~
I_n=
\begin{bmatrix}
1 &  0  \\
0 &  1
\end{bmatrix}\,.
\end{equation}
If the quadrupoles are tilted by 45$^\circ$ the 4$\times$4 transfer matrix can be written as
\begin{equation}
\label{simple_decoupling_matrix}
\overline R=R_r R_q R_r^T=\frac{1}{2}
\begin{bmatrix}
T_{n+} &  T_{n-}\\
T_{n-} &  T_{n+}
\end{bmatrix},
\end{equation}
where
\begin{equation}
R_r=\frac{\sqrt2}{2}
\begin{bmatrix}
I_n &  I_n  \\
-I_n &  I_n
\end{bmatrix}\,,\,\,\,\,
{T_{n\pm}=T_n \pm I_n}.
\end{equation}
The beam matrix $C_{3}^{'}$ after the decoupling section is
\begin{equation}
C_{3}^{'}=\overline R C_{2}^{'} {\overline R}^T=
\begin{bmatrix}
\eta_+ \Gamma_{n+} &  \zeta \Gamma_{n-} \\
\zeta \Gamma_{n-} &  \eta_- \Gamma_{n+}
\end{bmatrix},
\end{equation}
and the 2$\times$2 sub-matrices $\Gamma_{n\pm}$ are defined through
\begin{equation}
\Gamma_{n\pm}=
\begin{bmatrix}
U &  0  \\
0 &  \pm \frac{1}{U}
\end{bmatrix},
\end{equation}
with
\begin{align}
\label{eq2}
\eta_{\pm}=\frac{\varepsilon_n}{2}(\frac{\beta_n}{U}+\frac{U}{\beta_n} \pm 4k\beta_n),
\end{align}
and
\begin{align}
\label{eq2}
\zeta=\frac{\varepsilon_n}{2} (-\frac{\beta_n}{U}+\frac{U}{\beta_n})\,.
\end{align}
Assuming that this beam matrix is diagonal, its $x$-$y$ component vanishes
\begin{align}
\label{eq2}
\zeta \Gamma_{n-}=O_n\,.
\end{align}
This equation is solved by
\begin{align}
\label{eq2}
U=\beta_n\,.
\end{align}
This result was found earlier in~\cite{Kim} for instance. However, the major steps have been repeated here since they will be referred to later. 
\\
Suppose that the decoupling transfer matrix $\overline R$ is able to decouple the two transverse planes of $C_{2}^{'}$. We still do not know how this transfer beam line looks in detail,
but anyway we calculate the final rms emittances obtaining
\begin{align}
\label{eq2}
\varepsilon_{x,y}=\frac{\varepsilon_n}{2}(\frac{\beta_n}{U}+\frac{U}{\beta_n} \pm 4k\beta_n)\,.
\end{align}
This idealized example serves illustrating the principle, and it may be accomplished with just three skew quadrupoles. In our design, more elements are used because of finite apertures and gradients of a real experiment. In our set-up the decoupling section comprises a quadrupole triplet and a skew quadrupole triplet separated by a drift. The quadrupole gradients are optimized numerically from a numerical routine~\cite{Groening} to remove the inter-plane correlations thus minimizing the horizontal (for instance) rms emittances to the lower of the eigen-emittances. 
\\
Fig.~\ref{emittance transfer 1} illustrates the transverse emittance transfer. In the first step we assume that we turn off the power supplies of the solenoid and the skew quadrupole triplet. This process is an ordinary stripping process and the eigen-emittances are equal to the rms emittances at the exit of this section. It reflects today's situation of providing highly charged ions from linacs. Due to the stripping, growth of eigen-emittances and rms emittances is unavoidable. It is the reference scenario to which the emittance transfer scenario is to be compared. In the latter the solenoid field and the decoupling skew quads are turned on.
The eigen-emittances diverge inside the solenoid but they are preserved afterwards. Along the decoupling skew quadrupole triplet the rms emittances are made equal to the diverged
eigen-emittances. Compared to the reference scenario, the final horizontal rms emittance
is reduced significantly by a factor two. This emittance transfer experiment (EMTEX) is therefore fundamentally
different from an emittance exchange experiment (EEX). EMTEX is non-symplectic and the amount of transfer can be controlled by the solenoid field strength and/or the beam size on the stripping foil.
\\
\\
Behind the decoupling section another quadrupole triplet is required to rematch the beam for further transport to the SIS-18 synchrotron. The beam rms sizes along the total beam line are shown in~Fig.~\ref{beam rms sizes}
(solenoid and skew quads on) and the particle distributions at the exit of beam line are illustrated in~Fig.~\ref{particle distribution}.
\\
\section{Error Studies}
\begin{figure}[hbt]
\centering
\includegraphics*[width=80mm,clip=]{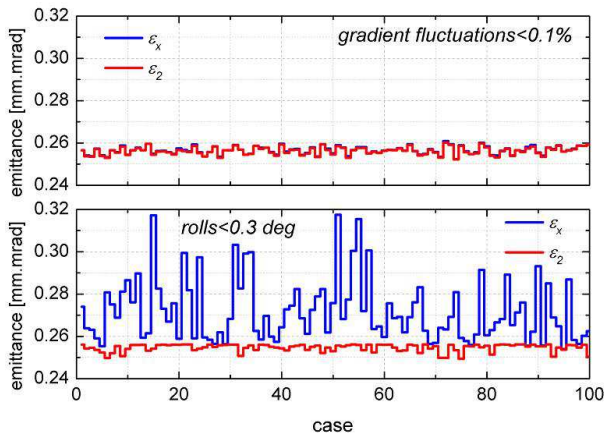}
\caption{Influence of gradient fluctuations and magnet rolls on the horizontal rms emittance and lower eigen-emittances.}
\label{influences of rotational errors 1}
\end{figure}

The decoupling of the transverse planes is sensitive to machine errors.
Alignment failures and/or gradient errors directly enter into the transformations and flaw the decoupling performance. In order to quantify the impact of such errors
on the experiment, dedicated error studies were done w.r.t. gradient errors and
rolls of quadrupoles. Gradient deviations and rolls were distributed randomly
among all magnets of the set-up. Different error distributions were used
to simulate the experiment by tracking. The influence of gradient fluctuations and rolls on the final horizontal rms emittance and lower eigen-emittances are shown in~Fig.~\ref{influences of rotational errors 1}. Based on the experiences of the UNILAC,
the maximum values of gradient fluctuations and rolls can be kept below 0.1$\%$ and 0.3$^\circ$, respectively.
\\
\\
It turned out that the gradient fluctuations are not critical. In case of rolls, the values of horizontal rms emittances are larger than the lower eigen-emittances, indicating incomplete decoupling. The related beam line errors degrade the accuracy of decoupling.
Assuming that the gradient fluctuations and rotational angles are lower than $\pm 0.1\%$ and 0.3$^\circ$, it does just slightly harm the decoupling capability, i.e. the coupling elements are removed sufficiently well.
\section{ Decoupling Capability Analysis}
\begin{figure}[hbt]
\centering
\includegraphics*[width=80mm,clip=]{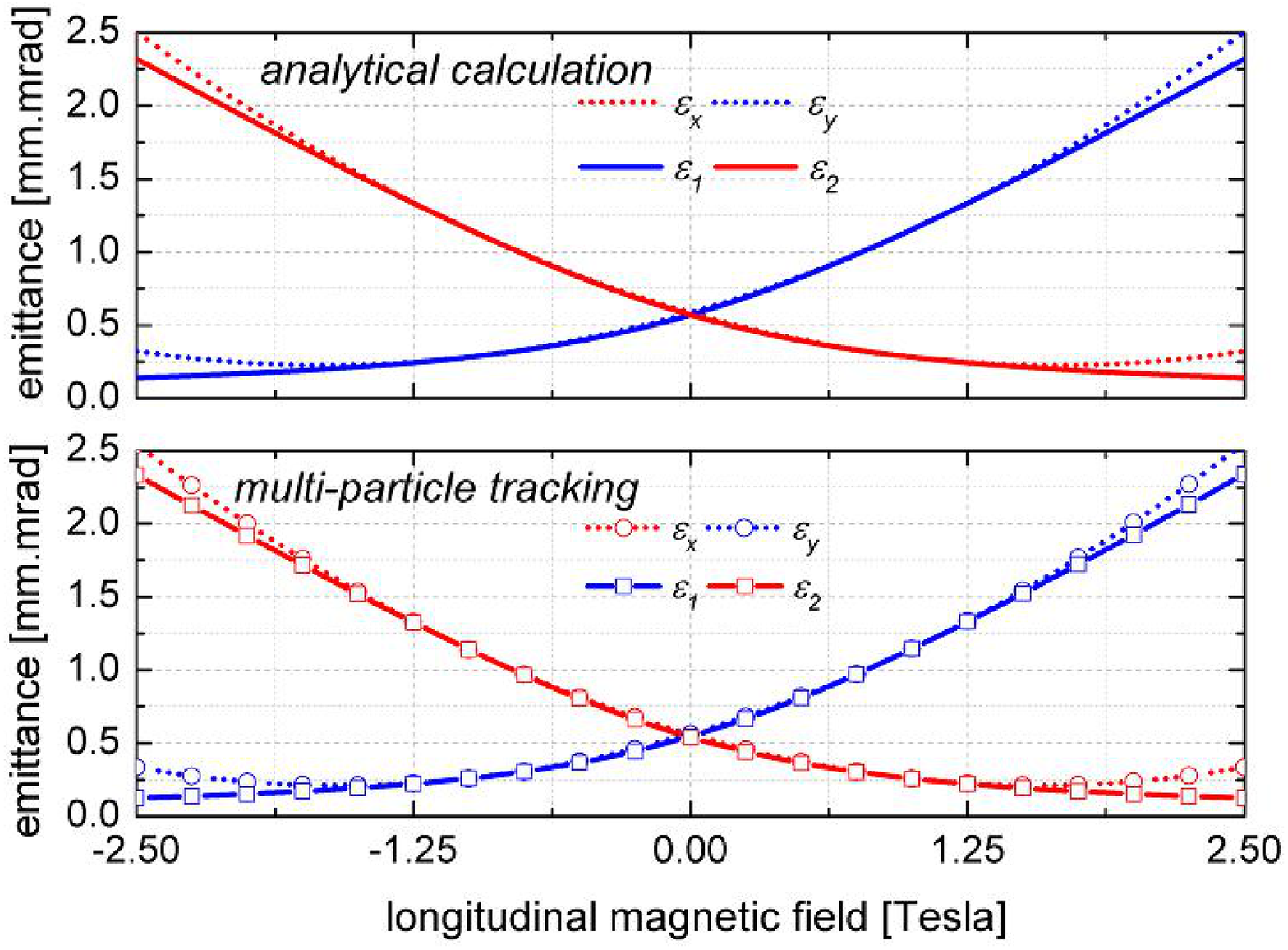}
\caption{Eigen-emittances and rms emittances calculated by analytical method based on the decoupling matrix of Eq.~(\ref{simple_decoupling_matrix}) and by multi-particle tracking through our specific set-up. Although the longitudinal magnetic field is varied, the decoupling gradients are kept constant at the values determined to decouple the beam coupled by a longitudinal magnetic field of 1~T.}
\label{analytical method}
\end{figure}

\begin{figure}[hbt]
\centering
\includegraphics*[width=80mm,clip=]{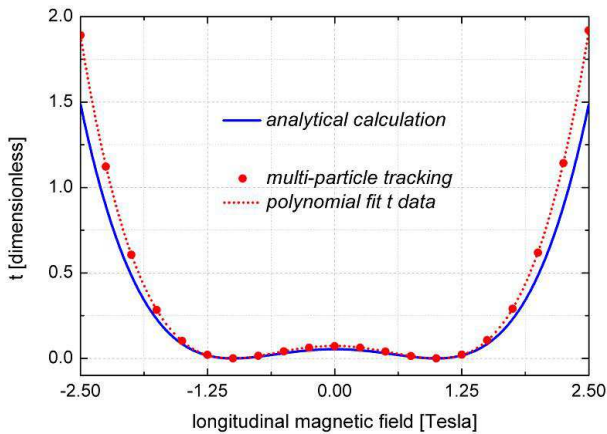}
\caption{The parameter $t$ calculated by analytical method and multi-particle tracking simulation. Although the longitudinal magnetic field is varied, the
decoupling gradients are kept constant at the values determined to decouple the
beam coupled by a longitudinal magnetic field of 1~T.}
\label{t}
\end{figure}

\begin{figure}[hbt]
\centering
\includegraphics*[width=80mm,clip=]{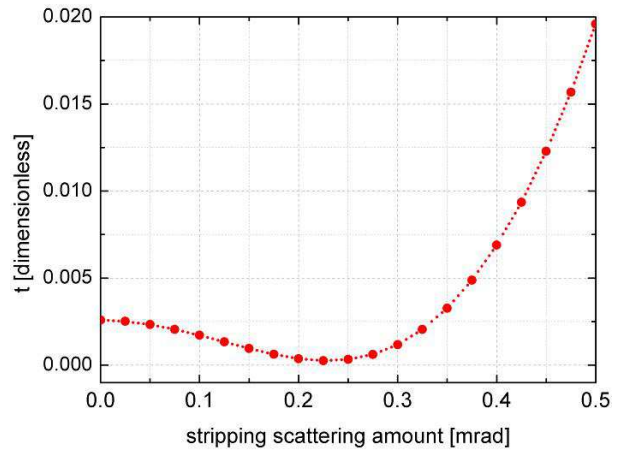}
\caption{The parameter $t$ calculated by multi-particle tracking simulation as a function of the foil scattering angle. Although the scattering angle is varied, the decoupling gradients are kept constant at the values
determined to decouple the beam coupled by a scattering angle of 0.226~mrad.}
\label{t2}
\end{figure}

The parameter $t$ is introduced to quantify the inter-plane coupling. If $t$
defined as
\begin{align}
\label{eq2}
t=\frac{\varepsilon_x \varepsilon_y}{\varepsilon_1 \varepsilon_2}-1
\end{align}
is equal to zero, there are no inter-plane correlations and the beam is fully decoupled. Kim~\cite{Kim} introduced the beam angular momentum $2\xi$=$\langle xy'-x'y\rangle $ and for an
angular momentum dominated beam one finds $t$=$\xi^2/\varepsilon_{4d}$. For a given solenoid strength $k_0$, referring to the unstripped beam, the corresponding quadrupole gradients of the decoupling section are determined using a numerical routine, such that finally the rms emittances are equal to the eigen-emittances. If these optimized gradients are applied to remove inter-plane correlations
produced by a different solenoid strength $k_1$, the resulting rms emittances and eigen-emittances at the exit of the decoupling section are calculated to be
\begin{align}
\label{eq2}
\varepsilon_{x,y}=\frac{\varepsilon_n(k_1)}{2}\left[\frac{\beta_n(k_1)}{\beta_n(k_0)}+\frac{\beta_n(k_0)}{\beta_n(k_1)} \pm  4k_1 \beta_n(k_1)\right],
\end{align}
and
\begin{align}
\label{eq2}
\varepsilon_{1,2}=\varepsilon_n(k_1)\left[  1 \pm 2k_1 \beta_n(k_1) \right].
\end{align}
The parameter $t$ is then
\begin{align}
\label{eq2}
t=\frac{ 4 \varepsilon^2 \beta^2  }{(\frac{\varepsilon}{\beta}+\Delta\varphi^2)(\frac{\varepsilon}{\beta}+4k_0^2 \varepsilon \beta+\Delta\varphi^2)}(k_1^2-k_0^2)^2\,.
\end{align}
\\
In the experiment, we will have a beam of molecules from $D_6^+$ with the initial beam parameters $\alpha$=0, $\beta$=2.5~mm/mrad and $\varepsilon$=0.51~mm.mrad at the entrance of the solenoid. The stripping scattering amount $\Delta \varphi$ is~0.226~mrad~\cite{Atima} and the decoupling transfer matrix is determined for 1.0~T of solenoid field.
For the simplest decoupling transfer matrix, the decoupling section is a skew quadrupole triplet. In our case, the decoupling section comprises
a quadrupole triplet and a skew quadrupole triplet separated by a drift. Therefore, our decoupling transfer matrix has a more complex structure, explicitly (in units of mm~mrad)
\begin{equation}
\overline R^{'}=
\begin{bmatrix}
-0.9224& 1.6051  & -0.4133 & -0.4703 \\
-0.7274&  0.2415  &  -1.6969 & -2.0649\\
0.0746 &  -0.2830 &  2.9308 & 3.6770 \\
0.4603 &  -1.0047 & 1.1329 & 1.7437
\end{bmatrix},
\end{equation}
being different from the form used in Eq.~(\ref{simple_decoupling_matrix}). The final eigen-emittances and rms emittances calculated using Eq.~(\ref{simple_decoupling_matrix}) and those obtained from tracking through our specific set-up are compared in~Fig.~\ref{analytical method}.
For the simple decoupling section the calculation is based on the transfer matrix method of Eq.~(\ref{simple_decoupling_matrix}). For our decoupling section multi-particle tracking through the external three-dimensional field maps (for the solenoid) and the external one-dimensional field profile (for the quadrupole and skew quadrupole) were adopted.
\\
The remarkable result is that both decoupling matrices work effectively for a wide range of longitudinal magnetic field values, i.e. the beam is well decoupled for a wide range of longitudinal magnetic fields around the gradients of the decoupling section the quadrupoles have been optimized for. Additionally, in both cases the decoupling performance is independent from the sign of $k_1$ as suggested by Eq.~(\ref{eq2}) and weakly depended on $k_1$-$k_0$.
We currently do not have a complete analytical understanding of this weak dependence except for the simple decoupling matrix Eq (\ref{simple_decoupling_matrix}). However, we still aim for understanding why the dependence is so weak even for our decoupling line being more complex w.r.t. the form of Eq.~(\ref{simple_decoupling_matrix}). To exclude that this is casual for this one beam line, the beam line has been modified by
prolonging or shortening drifts and quadrupole field lengths. For all modifications (all using a regular quadrupole triplet followed by a skew quadrupole triplet) the same behavior of the decoupling performance was observed.
\\
However, this behavior simplifies the
decoupling significantly as re-adoption of gradients to the solenoid field can be skipped within a reasonable range of solenoid fields. It provides an one-knob set-up to partition the horizontal and vertical beam rms emittances.
The behavior of $t$ calculated by the analytical method based on Eq.~(\ref{simple_decoupling_matrix}) and on tracking through our set-up is illustrated in~Fig.~\ref{t}, where the stripping scattering amount $\Delta \varphi$ is set to 0.226~mrad and the longitudinal magnetic field is varied.
In our set-up $k_0$ corresponds to a solenoid field of 1~T and accordingly $t$ has a
minimum for that value. The beam is well decoupled for a wide range of solenoid fields for both the analytical calculation and for tracking through the specific set-up.
\\
\\
The dependence of $t$ on the solenoid field as obtained from tracking has been fitted with a 4th order polyoma as motivated by Eq.~(\ref{eq2}) and the fit is plotted as well in~Fig.~\ref{t}. This result might suggest a general 4th order dependence of the
decoupling performance of any beam line on the coupling-driving solenoid field. The
analytical investigation of this suggestion is beyond the scope of this paper.
\\
Finally we calculated the case of fixed longitudinal magnetic field but different stripping scattering amount $\Delta \varphi$, and the behavior of $t$ simulated by multi-particle tracking through our set-up is illustrated in~Fig.~\ref{t2}. The decoupling is also quite independent from the amount of scattering, which will additionally facilitate the experiment.
\\
\\
\section{Conclusion and Outlook}
\begin{figure}[hbt]
\centering
\includegraphics*[width=30mm,clip=]{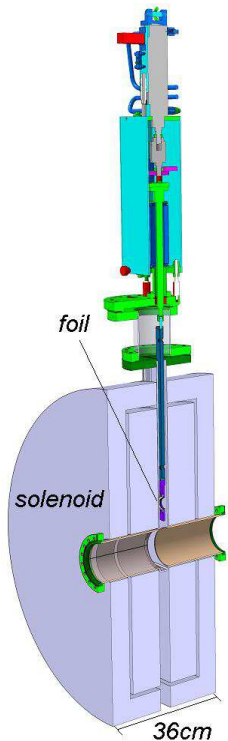}
\caption{The solenoid comprising to coils, its chamber, and the driver mechanics to place the stripping foil on the beam axis.}
\label{solSsrip}
\end{figure}

An experimental set-up for demonstration of round-to-flat transformation of an initially
decoupled ion beam was presented. It comprises two doublets for matching the required
beam parameters on a stripping foil being placed in the center of a solenoid of
about 1.0~T. The net effect on the beam is a non-symplectic transformation
creating inter-plane coupling, being removed afterwards along a beam line from
one regular quadrupole triplet and one skew quadrupole triplet. Extensive tracking simulations through
three-dimensional field maps of the solenoid were performed for a variety of field shapes, showing excellent
agreement to the pure matrix formalism. Angular
scattering during stripping was included and an error study was performed. The latter
revealed that quadrupole rolls may slightly but not significantly harm the decoupling
performance. This decoupling performance was found to be very stable w.r.t. the field
strength of the solenoid, i.e. the same decoupling gradients can be applied for a
wide range of solenoid fields without relevant reduction of the decoupling performance.
This remarkable result can be partially understood analytically. Apart from that it
facilitates the conduction of the experiment itself.
\\
The beam line is currently under construction. Quadrupole triplets and doublets are on site or under production. Power converters were ordered and the
installation of the infrastructure is scheduled. The solenoid, comprising two separate coils, including its chamber, which in turn
houses the driver to move the stripping foil on the beam axis, is shown in~Fig~\ref{solSsrip}. The solenoid design avoids a local field
minimum in the center, since it might act as a trap of electrons.
All required diagnostic devices are already installed and operational. We currently plan to do the experiment in 2014, that work is supported by the HIC for FAIR and the BMBF.
\\
\begin{acknowledgments}
The authors wish to express their gratitude to Brahim Mustapha at ANL, and Ina Pschorn at GSI for fruitful discussions.
\end{acknowledgments}

\appendix
\section{}
The beam parameters at the entrance and exit of the beam line are listed in~Tab.~\ref{tab_1}. A $D_6^+$ beam of 11.4 MeV/u is stripped in a foil to a 3$D_2^+$ beam. The total relative momentum spread is less than $\pm5 \times 10^{-4}$.
The parameters of the full beam line are listed in~Tab.~\ref{tab_2}. Positive gradient means horizontal focusing and a skew refer to a normal quadrupole rotated counter-clockwise by 45$^\circ$ around the beam line axis.
\\
\begin{table}
\caption{\label{tab_1}The beam parameters at the entrance and exit of the EMTEX beam line.}
\begin{ruledtabular}
\begin{tabular}{c|cc}
Parameters & Entrance & Exit \\
\hline
$\alpha_x$/$\alpha_y$ & -1.21/-2.28 &0.00/0.00 \\
$\beta_x$/$\beta_y$ [mm/mrad] & 21.80/15.01 &7.18/7.13 \\
$\varepsilon_x$/$\varepsilon_y$ [mm.mrad] & 0.509/0.510 &0.256/1.144
\end{tabular}
\end{ruledtabular}
\end{table}
\\
\begin{table}
\caption{\label{tab_2}The lattice of the EMTEX beam line. }
\begin{ruledtabular}
\begin{tabular}{c|cc}
Element & Effective Length [mm] & Gradient [Tesla/m]\\
\hline
Drift & 240.5 &  \\
Quad & 319.0 & 7.276\\
Drift & 203.0 &\\
Quad& 319.0 & -7.726\\
Drift & 4000.0 &\\
Quad & 354.0 & -0.187\\
Drift & 167.5 &\\
Quad & 354.0 & 3.287\\
Drift & 500.0 &\\
\hline
Drift & 300.0 & \\
Solenoid & 100.0 & 1.00 Tesla \\
Foil & 0.0 &     20~$\mu g/cm^2$, $\Delta \varphi$=0.226~mrad \\
Solenoid & 100.0 & 1.00 Tesla \\
Drift & 300.0 & \\
\hline
Drift & 200.0 &\\
Quad & 319.0 & 10.600\\
Drift & 201.0 &\\
Quad & 319.0 & -9.453\\
Drift & 201.0 &\\
Quad & 319.0 & 8.386\\
Drift & 500.0 &\\
Skew Quad & 200.0 & -5.618\\
Drift & 20.0 &\\
Skew Quad & 400.0 & 2.840\\
Drift & 20.0 &\\
Skew Quad & 200.0 & -9.106\\
\hline
Drift & 500.0 &\\
Quad & 200.0 & -6.813\\
Drift & 20.0 &\\
Quad & 400.0 & 7.356\\
Drift & 20.0 &\\
Quad & 200.0 & -7.760\\
Drift & 1289.0 &
\end{tabular}
\end{ruledtabular}
\end{table}
\\

\bibliography{C.Xiao}

\begin{thebibliography}{20}
\bibitem{Non-symplecticity}
The non-symplecticity is from omission of parts of the full system comprising the stripping process. It includes the incoming beam particle nuclei, their residual electrons, and the nulei and electrons
of the stripping foil atoms. However, for the beam dynamics just the stripped beam ions are kept in the system. The stripping atoms and the stripped-off electrons are removed artificially from the system. This removal is non-symplectic.
\end{thebibliography}
\end{document}